\documentclass[reprint,letterpaper,twocolumn,superscriptaddress,amssymb, nobibnotes, aps, pra, showpacs]{revtex4-1}
\usepackage %[dvipdfm] this is for picture in JPEG/PNG/BMP file?
{graphicx,amsmath,amssymb}

\begin{document}

\title{Observation of Rapid Change of Crystalline Structure during the Phase Transition of the Palladium-Hydrogen System}

\author{Akio Kawasaki}%
\email{akiok@mit.edu}
\affiliation{Department of Physics, Massachusetts Institute of Technology, Cambridge, Massachusetts 02139, USA}
\affiliation{Department of Physics, University of Tokyo, Bunkyo, Tokyo 113-0033, Japan}
\affiliation{RIKEN, Nishina Center, Wako, Saitama-ken 351-0198, Japan}

\author{Satoshi Itoh}%
\affiliation{Department of Physics, University of Tokyo, Bunkyo, Tokyo 113-0033, Japan}
\affiliation{RIKEN, Nishina Center, Wako, Saitama-ken 351-0198, Japan}

\author{Kunihiro Shima}
\affiliation{Tanaka Kikinzoku Kogyo K.K., Tomioka, Gunma-ken 370-2452, Japan }

\author{Kenichi Kato}%
\affiliation{RIKEN SPring-8 Center, Sayo, Hyogo-ken, 679-5148, Japan}

\author{Haruhiko Ohashi}%
\affiliation{RIKEN SPring-8 Center, Sayo, Hyogo-ken, 679-5148, Japan}

\author{Tetsuya Ishikawa}%
\affiliation{RIKEN SPring-8 Center, Sayo, Hyogo-ken, 679-5148, Japan}

\author{Toshimitsu Yamazaki}%
\affiliation{Department of Physics, University of Tokyo, Bunkyo, Tokyo 113-0033, Japan}
\affiliation{RIKEN, Nishina Center, Wako, Saitama-ken 351-0198, Japan}

%\texttt{\jobname.tex}
\date{\today}

\begin{abstract}
We performed an X-ray diffraction experiment while palladium bulk absorbed and desorbed hydrogen to investigate the behavior of the crystalline lattice during the phase transition between the $\alpha$ phase and $\beta$ phase. Fast growth of $\beta$ phase was observed around $x=0.1$ and $x=0.45$ of PdH$_{\rm x}$. In addition, slight compression of the lattice at high hydrogen concentration and increase in the lattice constant and the line width of the $\alpha$ phase after a cycle of absorption and desorption of hydrogen was observed. These behavior correlated with the change in the sample length, which may infer that the change in shape was related to the phase transition. 
\end{abstract}

\pacs{61.50.Ks,64.70.K-,64.70.kd, 81.05.Bx}

\maketitle
\section{Introduction}
Palladium is known as a metal that absorbs large amount of hydrogen. This property opened its applications to storage and filter of hydrogen, and intense research on this property has been performed \cite{PdHSystem,AnnRevMatSci.21.269}. 
The large amount of hydrogen absorption is related to two phases of the palladium-hydrogen (Pd-H) system, one called $\alpha$ phase and the other called $\beta$ phase. The $\alpha$ phase has smaller hydrogen fraction $x={\rm H/Pd}$, and it includes pure palladium. The $\beta$ phase contains more hydrogen atoms. When palladium metal is exposed to hydrogen gas, typically at high temperature such as 100 $^{\circ}$C or more, palladium absorbs hydrogen. The absorption induces phase transition from $\alpha$ phase to $\beta$ phase through $\alpha +\beta$ phase, where the $\alpha$ phase and the $\beta$ phase coexist, as the phase diagram \cite{PlatMetRev.21.44,HydrogenMetalsII,MatSciEngA.551.231} shows. 

This phase transition from the $\alpha$ phase to the $\beta$ phase is believed to be the cause of deformation of a palladium bulk reported in Refs. \cite{TransElectrochemSoc.68.449, PlatinMetalsRev.4.130, SurfTechnol.16.57, IntJHydrogenEnergy.22.175, ActaMater.46.4543, JSocMatSciJpn.49.1242, JSocMatSciJpn.50.999,ProcJpnAcadSerB.85.183,MatSciEngA.551.231}. These changes in the shape significantly larger than the ordinary plastic deformation are explained as a phenomenon connected to a phase transition, sometimes with a relation to superplasticity. 
We previously found a change in the shape of palladium metal in the direction of minimizing its surface area \cite{ProcJpnAcadSerB.85.183}, and subsequently we observed large bending of a horizontal palladium plate with only a small external force, and warping back and forth of vertical palladium plate \cite{MatSciEngA.551.231}. However, it is not clear why the change in the shape of the bulk happened so as to minimize its surface area and why the palladium plate warped back and forth. 
\begin{figure}[!t]
 \begin{center}
 \includegraphics[width=0.6\columnwidth]{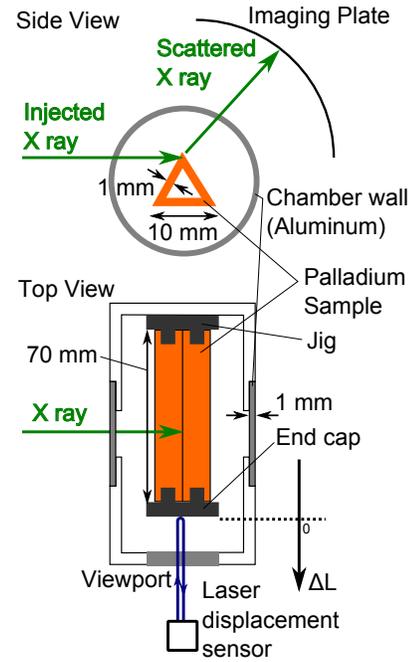}
 \caption{(Color online) Experiment Setup} 
 \label{Setup}
 \end{center}
\end{figure}

In this paper, we report an X-ray diffraction experiment at a synchrotron radiation facility SPring-8 to obtain microscopic information of the Pd-H system during the phase transition. We took X-ray diffraction spectra while a palladium bulk absorbed and desorbed hydrogen gas with the information on the length of the sample. The analysis was performed to get microscopic information, such as lattice constant, crystal grain size, and the intensity of the diffraction from the two phases, against the time and the hydrogen fraction $x$, in order for us to obtain microscopic understanding of the change in the shape of the palladium bulk during the treatment with hydrogen.  

Microscopic study of the Pd-H system has long been performed \cite{ProcPhysSoc.49.587,AnnRevMatSci.21.269,TransFaradaySoc.58.1989,JLessCommonMetal.88.115}.  Initially, X-ray diffraction study was performed in the static states. The lattice constants of the $\alpha$ phase and the $\beta$ phase are measured as $a_\alpha=3.894$ \AA \ and $a_\beta=4.040$ \AA \ \cite{AnnRevMatSci.21.269} by the standard X-ray diffraction technique after putting the sample back to the room temperature. The figures in Ref. \cite{TransFaradaySoc.58.1989} was obtained by placing X-ray diffraction pictures of different samples with different hydrogen fraction together. The dynamic observation of the phase transition has also been performed \cite{NatMater.13.802,JCatalysis.70.298,IntJHydrogenEnergy.35.8609,JPhysChemB.105.8088}. Most of them use the information of a single peak to estimate the amount of one phase in the bulk. Lattice constant information on these reports is simply plotted against temperature or pressure of hydrogen, and time dependent analysis was not performed. In addition, most of the reports are for nano particle of palladium \cite{JCatalysis.70.298,IntJHydrogenEnergy.35.8609,JPhysChemB.105.8088}, and the time dependent microscopic study of the palladium bulk is lacking. Our report fills this blank and connects the relation between the macroscopic behavior and the microscopic parameter.

\section{Experimental Method}
\subsection{Experiment setup}
The X-ray diffraction experiment was performed at RIKEN Materials Science beamline BL44B2 of SPring-8. The X-ray energy was 20 keV with the energy resolution of $10^{-4}$. The photon flux was $10^{11}$ s$^{-1}$ and the beam size was 0.5 mm (vertical) $\times$ 3.0 mm (horizontal). The palladium sample was 70 mm tall, 1 mm thick hollow equilateral triangular tube with 10 mm wide outer side. This shape ensured high mechanical strength to prevent the sample from bending. The top corner of the sample was aligned to the path of incident X-ray beam.

The palladium sample was put into a vacuum chamber with one end fixed onto the chamber. The other end was free and with a flat end cap, being enabled to change the length. The sample length change $\Delta L$ was measured with a laser displacement sensor through a viewport. The chamber was made of stainless steel, except for the path for the X-ray made of 1 mm thick aluminum. The chamber was connected to a pumping system consisting of a turbo molecular pump and a rotary pump. The whole chamber was wrapped with heating tapes to heat the sample up to 120 $^{\circ}$C. 

Prior to the start of the experiment, the chamber was evacuated out to below $10^{-4}$ Pa. Hydrogen gas was introduced through a flow meter, the start of which is defined as $t=0$. At first, the introduction was at a maximum rate of 40 ml/min, and once the pressure reached 0.2 MPa, the flow rate was reduced in order to keep the pressure around 0.2 MPa, as shown in Fig. \ref{AlphaAngle} (e). After the saturation of the pressure and $\Delta L$, by which we regarded the phase transition as finished, we pumped the hydrogen out through the pumping system. We finished the evacuation when the decrease in pressure accelerated, which is the sign of the complete outgassing of hydrogen. We introduced hydrogen again to see the behavior at the beginning of the second cycle of absorption/desorption of hydrogen. Throughout the experiment, we recorded the pressure and the temperature of the chamber and $\Delta L$. 

The scattered X-ray was recorded with an imaging plate covering the diffraction angle $2\theta$ of $0\leq 2\theta \leq 78^{\circ}$. One imaging plate recorded data of 18 different exposures and therefore one set of measurements consisted of 18 data. Each exposure was for 30 seconds. We started the first exposure of the first set right after we started to introduce hydrogen. 

\subsection{Characteristics of the setup}
Although the result and its interpretation are described assuming the sample was in the ideal condition for the powder X-ray diffraction, 
we have to consider the difference between the ideal condition and our system. First, palladium sample was large and certain amount of the X-ray was absorbed by the sample. Second, the number of crystal grains might be small and their orientation might not be random. Third, the sample might not have been uniform, as it takes certain amount of time for hydrogen to diffuse into the depth of the palladium. 

The Attenuation coefficient of the palladium is 207.9 cm$^{-1}$ for 20 keV X-ray. This means that the X-ray passing through 0.1 mm bulk is attenuated to 13\% of the incoming flux. Thus, we basically looked at the phase transition of the thin surface layer whose thickness is at most 0.1 mm. As the X-ray hit the top point of the triangular cross section of the sample, we can assume that the reasonable amount of the diffracted light is transmitted for all the diffraction angles. 

The volume we observed is roughly 0.5 mm $\times$ 3 mm $\times$ 0.1 mm, whereas the grain size of palladium is 10-100 $\mu$m in Fig. \ref{SEM}.  
It is therefore possible that the number of grains is too small for the diffraction pattern to have circular symmetry.  This can cause the suppression of the signal from certain crystal planes.  
\begin{figure}[!tb]
 \begin{center}
 \includegraphics[width=0.8\columnwidth]{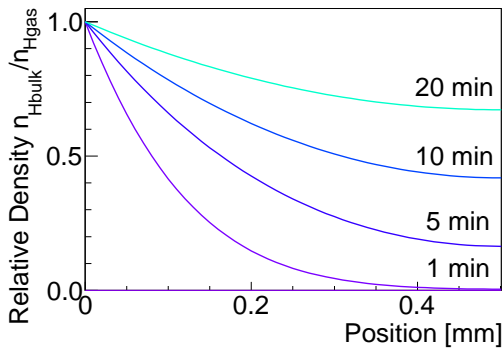}
 \caption{(Color online) Density of hydrogen in 1mm thick palladium bulk relative to the gas phase 1, 5, 10 and 20 minutes after the start of the introduction of hydrogen: initial condition is all point was zero at $t=0$ and the number at the two ends increased at a constant rate to simulate the increasing hydrogen pressure.} 
 \label{Diffusion}
 \end{center}
\end{figure}

Given that we looked at the thin surface layer, the diffusion would not take that long time. Fig. \ref{Diffusion} shows the numerical calculation of the absorption of hydrogen by 1 mm thick palladium plate when the density of the hydrogen gas linearly increased, with the diffusion coefficient of $3.26 \times 10^{-10}$ for hydrogen in palladium at 120 $^{\circ}$C. Although it took certain amount of time for hydrogen to reach the center of the sample, the 0.1 mm thick surface layer got the amount of hydrogen comparable to the very surface in short time; in 1 minutes, the density becomes 40\% of the outermost area and after 20 minutes, the density is 90\% of the surface one. Thus, except for the first few minutes, hydrogen density can be assumed to be roughly uniform.

\section{Results}
\begin{figure*}[!tb]
 \begin{center}
 \includegraphics[width=2.0\columnwidth]{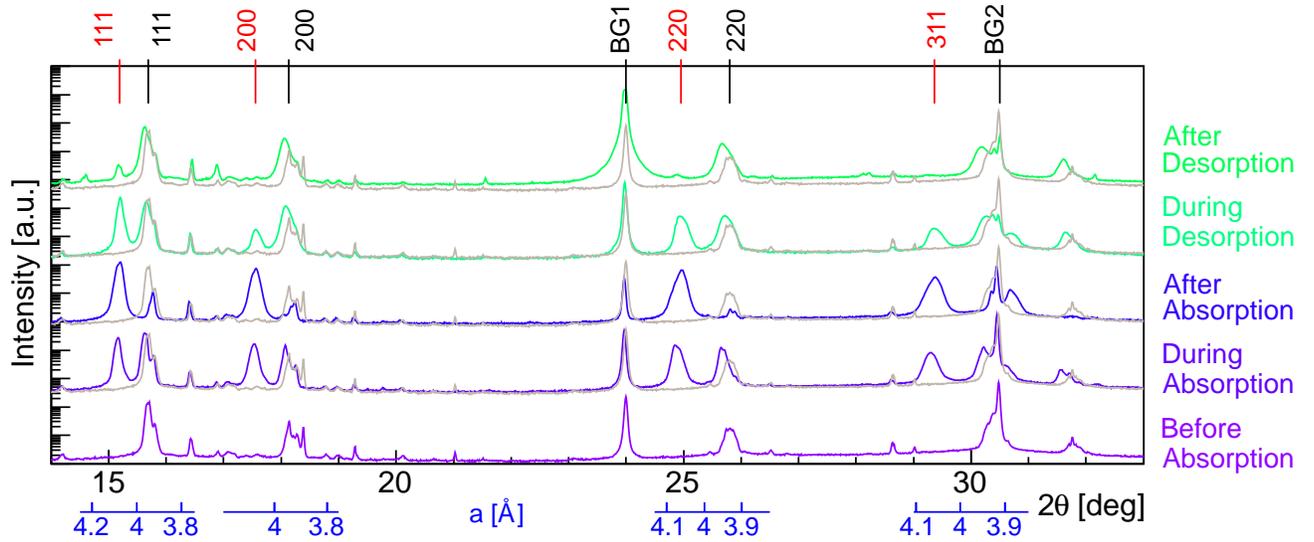}
 \caption{(Color online) X-ray diffraction spectra over experiment run: the middle of the absorption and the desorption spectrum was 12 minutes and 90 minutes after the start of the absorption and the desorption, respectively. Three-digit numbers on the top represent the crystal plane identification for the $\alpha$ phase (black) and the $\beta$ phase (red). BG1 and BG2 are background lines. Blue horizontal axes show the lattice constant for different crystal plane. The gray lines superposed onto the four spectra are the spectrum before absorption. }
 \label{No.0}
 \end{center}
\end{figure*}

Figure \ref{No.0} shows diffraction spectra at five representative moments during the experiment. Some conspicuous diffraction lines before the absorption of hydrogen got weaker during the absorption, and disappeared after the absorption. When palladium desorbed hydrogen, these peaks reappeared roughly at the same $2\theta$. During the absorption, new diffraction lines appeared at slightly smaller $2\theta$, which corresponds to the lattice constant $a$ larger than the vanishing diffraction lines. These new lines disappeared during the desorption. Based on the observation, we identified the phase and the crystal plane for the diffraction lines. The intense diffraction lines at the beginning were identified as $\alpha$ phase lines, as the sample was annealed before the experiment to outgas hydrogen. The new lines appearing during the absorption were identified as $\beta$ phase lines because these had slightly larger lattice constant $a_\beta$ than $a_\alpha$, which matched with the previously reported ratio of $a_\alpha$ and $a_\beta$ in Ref. \cite{AnnRevMatSci.21.269} (subscription $\alpha$ and $\beta$ show parameters for the $\alpha$ and the $\beta$ phase, respectively).  In addition, the difference between $a_\alpha$ and $a_\beta$ was significantly larger than the fluctuation of $a_\alpha$ and $a_\beta$ during the absorption and the desorption of hydrogen. Diffraction lines without any significant change over the experiment were regarded to be the background originated from other materials on the path of X-ray, such as the aluminum chamber wall and resistive material in the heating tape.  These lines were used to remove systematic errors of the diffraction angles.

\begin{figure}[!tb]
 \begin{center}
 \includegraphics[width=1.0\columnwidth]{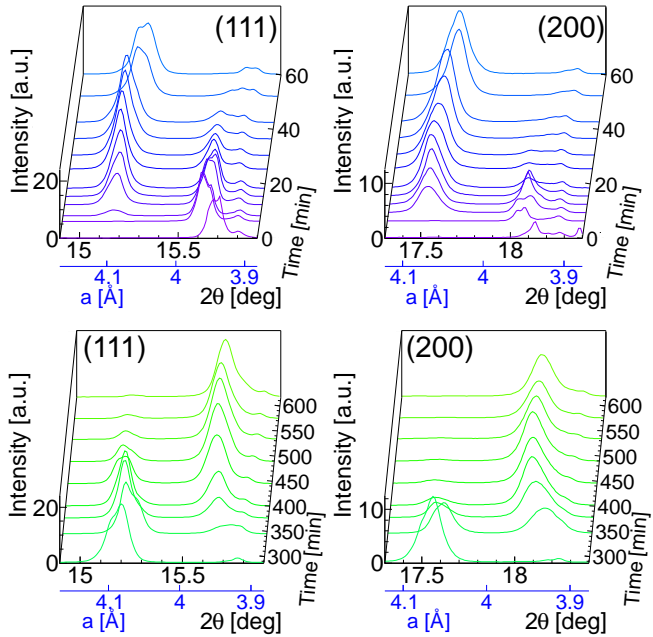}
 \caption{(Color online) Magnified X-ray diffraction spectrum of (111) line (top left: absorption stage, bottom left: desorption stage) and (200) line (top right: absorption stage, bottom right: desorption stage). The blue axes shows the lattice constant. }
 \label{MagAbs}
 \end{center}
\end{figure}

Figure \ref{MagAbs} is magnified plots of the time dependent behavior of the diffraction lines from the (111) and (200) planes.
The top half shows the first 60 minutes of the absorption stage. The initially intense $\alpha$ phase diffraction lines gradually disappeared over an hour and $\beta$ phase diffraction lines grew up rapidly at significantly smaller $2\theta$ after several minutes. It is notable that most of the growth of the $\beta$ phase diffraction lines finished in several minutes. The bottom half of Fig. \ref{MagAbs} shows the behavior during the desorption stage. The $\beta$ phase diffraction lines gradually disappeared and the $\alpha$ phase diffraction lines grew up. The diffraction line angle changed as time evolved. Changes were slower during the desorption stage than that during the absorption stage, but they had a common behavior that the growing peaks grew up smoothly, whereas the disappearing peaks had fluctuation in their position and the intensity during its fast vanish.  

\begin{figure*}[!tb]
 \begin{center}
 \includegraphics[width=2.0\columnwidth]{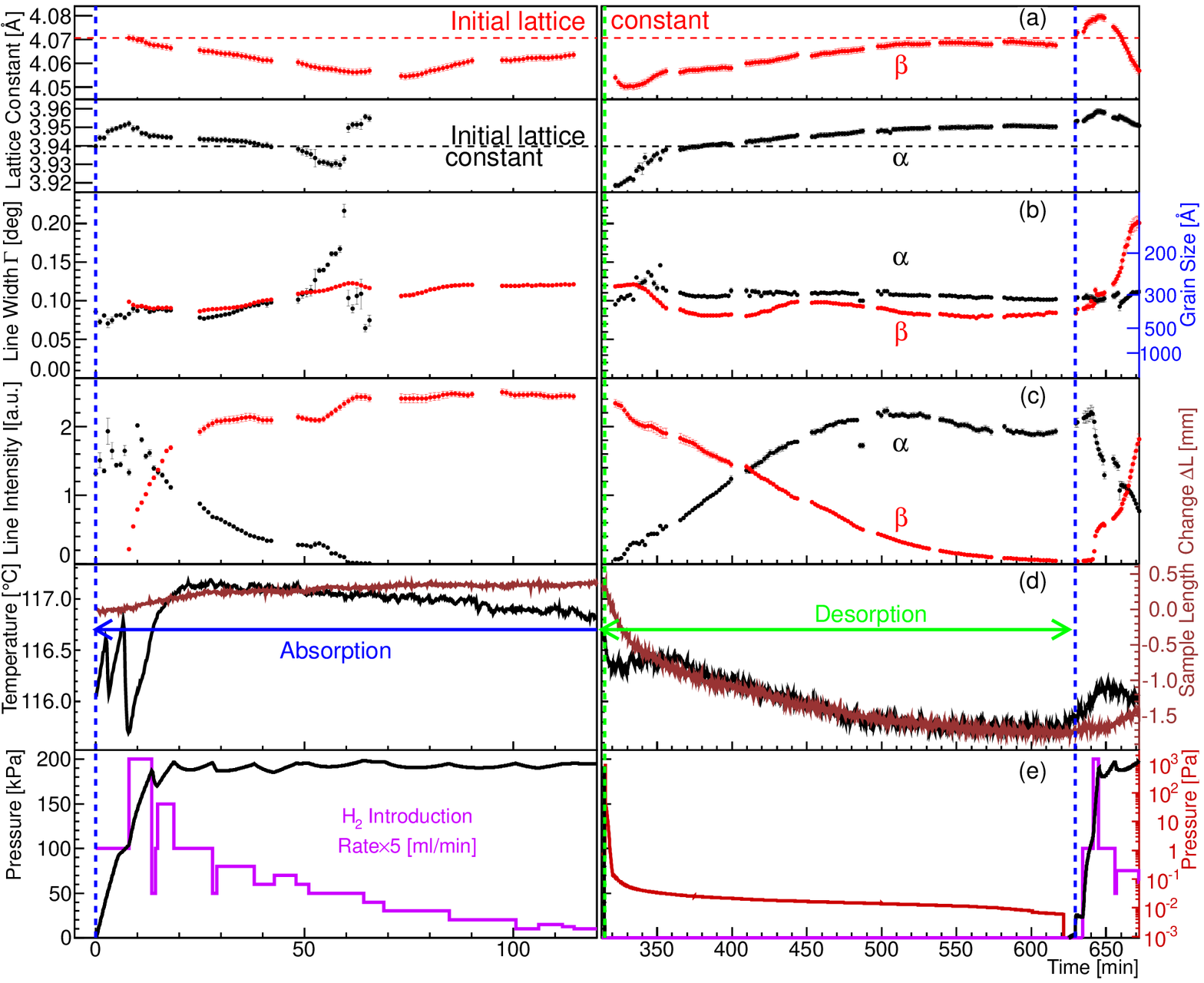}
 \caption{(Color online) Time dependence of (a) the lattice constant $a_\alpha$ and $a_\beta$, (b) line width $\Gamma_\alpha$ and $\Gamma_\beta$ and corresponding grain size (right axis), and (c) line intensity $I_{\alpha}$ and $I_{\beta}$ of the diffraction from (111) plane. Black points are for the $\alpha$ phase and red ones are for the $\beta$ phase. The sample length change $\Delta L$ is shown in (d) (brown line, positive $\Delta L$ corresponds to larger sample), and experiments condition are shown in (d) and (e). The H$_2$ introduction rate is magnified by 5 times, and should be read with the left axis. Blue and green dotted lines show the start of absorption and desorption stage of the hydrogen. Data points are not shown if the diffraction line was too weak to perform the fit. Only initial 120 minutes is shown for the first absorption stage, as the last 200 minutes did not have much change.} 
 \label{AlphaAngle}
 \end{center}
\end{figure*}

For a close look at the time dependent behavior of the diffraction line parameters, we fitted each diffraction line with a Gaussian plus a linear background function, obtaining the center of the diffraction line $2\theta_\alpha$ and $2\theta_\beta$, the diffraction line width $\Gamma_\alpha$ and $\Gamma_\beta$ as the full width at half maximum (FWHM) and the diffraction line intensity $I_\alpha$ and $I_\beta$ as the area of the Gaussian. If two or more diffraction lines were close to each other, the fitting function contained multiple Gaussian functions. Among these data, the (111) diffraction line is plotted in Fig. \ref{AlphaAngle} as the representative. Fig. \ref{AlphaAngle} (a) shows $a_\alpha$ and $a_\beta$ calculated from $2\theta_\alpha$ and $2\theta_\beta$. A systematic error of $2\theta_\alpha$ and $2\theta_\beta$ presumably due to the slight fluctuation of the position of the whole chamber against the beam and the imaging plate was removed by adding a correction so as to keep the diffraction angle of BG1 line in Fig. \ref{No.0} constant at the average value over all the data points.  Fig. \ref{AlphaAngle} (b) shows $\Gamma_\alpha$ and $\Gamma_\beta$ in the unit of angle. Fig. \ref{AlphaAngle} (c) displays $I_{\alpha}$ and $I_{\beta}$. A systematic fluctuation of $I_{\alpha}$ and $I_{\beta}$ was removed by a compensation factor that made an average of $I_{\beta}$ over the five last sets of absorption stage constant. 

$\Delta L$ is plotted in Fig. \ref{AlphaAngle} (d). The temperature of the sample, pressure inside the chamber and the hydrogen flow rate are plotted in Fig. \ref{AlphaAngle} (d) and (e). 

\begin{figure}[!tb]
 \begin{center}
 \includegraphics[width=0.95\columnwidth]{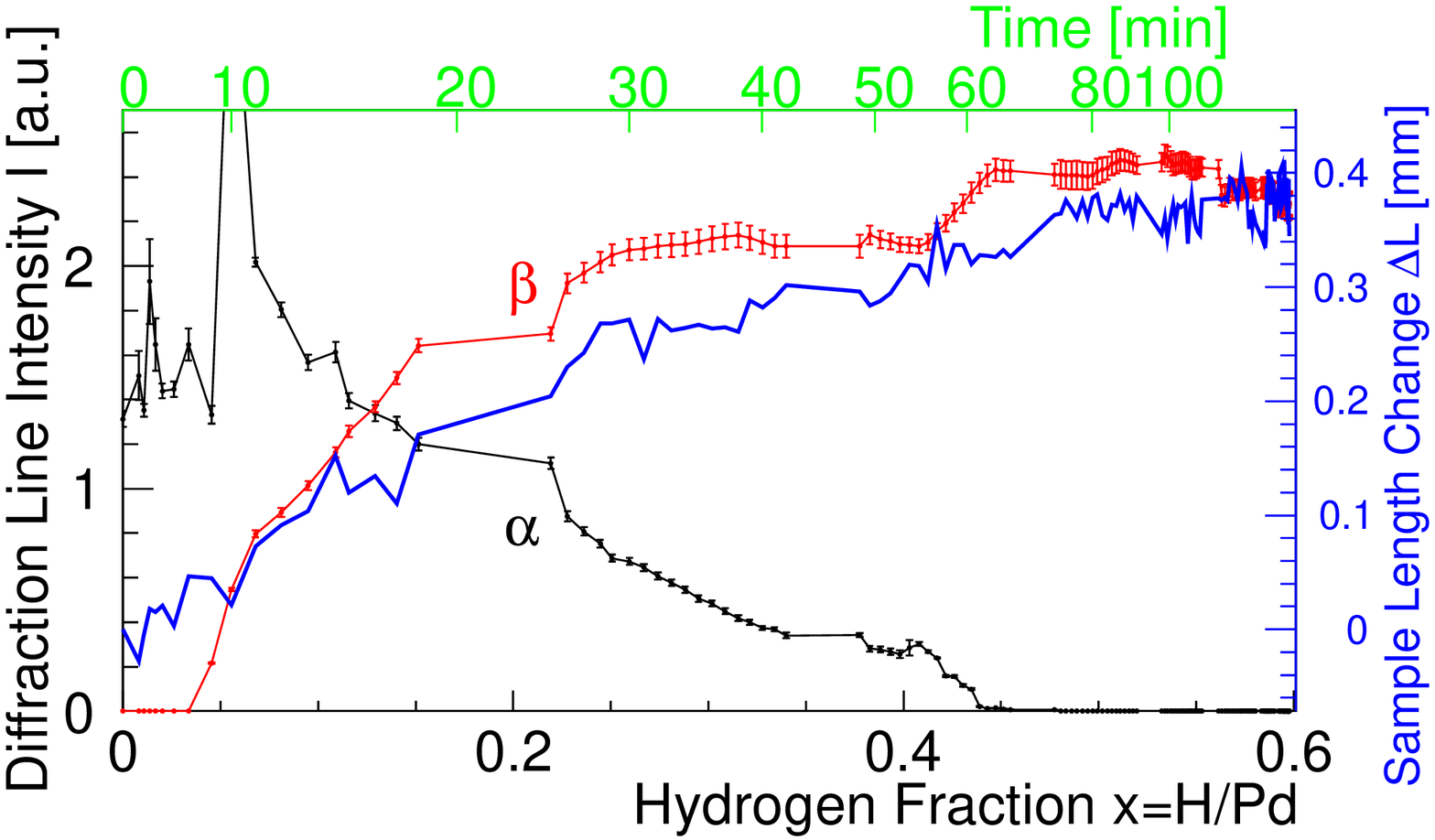}
 \caption{(Color online) Intensity of the (111) diffraction line against the hydrogen fraction $x$.} 
 \label{VSH}
 \end{center}
\end{figure}

Figure \ref{VSH} shows $I_{\alpha}$ and $I_{\beta}$ of (111) line against the hydrogen fraction $x$, together with $\Delta L$. The $x$ was calculated from the chamber volume, the hydrogen flow rate and the pressure.

\section{Discussion}
\subsection{Lattice constant $a_{\alpha}$ and $a_{\beta}$}
$a_{\alpha}$ and $a_{\beta}$ based on the spectrum before and after the absorption in Fig. \ref{No.0} are summarized in Table \ref{LatticeConstant}. Our result gave larger number than Ref. \cite{AnnRevMatSci.21.269}.  Part of the reason of the difference is the thermal expansion of the lattice. The thermal expansion coefficient for pure palladium, $1.18 \times 10^{-5}$ K$^{-1}$ at 293 K, and 95 K temperature difference (Ref. \cite{AnnRevMatSci.21.269}: 25 $^{\circ}$C, us: 120 $^{\circ}$C) gives 0.004 \AA \ expansion of the lattice. However, this still leaves discrepancy significantly larger than the statistical fluctuation, particularly for $a_{\alpha}$.  The remaining difference should be explained by the different condition to prepare the sample, as the difference is in the order of magnitude same as the fluctuation of $a_{\alpha}$ and $a_{\beta}$ shown in Fig. \ref{AlphaAngle}.

\begin{table}[!t]
 \caption{Lattice constant [\AA]: TE column shows the amount of the thermal expansion of the lattice.} 
\begin{tabular}{lcrclcccccc}
  \hline   \hline 
Phase& &\multicolumn{3}{c}{This result} &\hspace{2mm} & Ref. \cite{AnnRevMatSci.21.269} &\hspace{2mm} & Difference &\hspace{2mm} & TE \\
  \hline   
$\alpha$& &	3.928	& $\pm$ & $0.005 $	& & 3.894	& & 0.034 & & 0.004\\
$\beta$& &	4.054	& $\pm$ & $0.006 $	& & 4.040& & 0.014 & & 0.004\\
    \hline   \hline 
\end{tabular}
\label{LatticeConstant}
\end{table}

The time dependent behavior of $a_{\alpha}$ and $a_{\beta}$ of the (111) diffraction line is shown in Fig. \ref{AlphaAngle} (a). Notable features for $a_{\alpha}$ are (i) its initial increase, (ii) smaller $a_{\alpha}$ than the initial value when the $\alpha$ phase appeared again at the desorption stage, and  (iii) the increase of $a_{\alpha}$ at the end of the desorption. The first feature should be simply due to the hydrogen's occupying interstitial space among palladium atoms to expand the palladium lattice. 

$a_{\alpha}$ was 0.6\% smaller than the initial valuewhen it appeared again during the desorption. This is significantly larger than the uncertaintiy of the center position, and is possibly due to the compression by the $\beta$ phase. Because the sample length $L$ increased by only 0.6\% even when the sample completely turned into $\beta$ phase that has 3\% larger lattice constant than $\alpha$ phase, the crystal grain experienced the huge internal stress\cite{PhysChemChemPhys.13.11412}, and it is possible that the pressure compressed $\alpha$ phase lattice. The fact that the rapid growth of the $I_\beta$ started at the same time as the start of the decrease in $a_\alpha$ during absorptionstarted also supports the idea of the compression by the $\beta$ phase. 

$a_{\alpha}$ increased by 0.3\% after a cycle of the absorption and the desorption of the hydrogen, which is significantly larger than the uncertainty of the center position. This means that there is an irreversible effect of hydrogen treatment on palladium bulk. The behavior of $a_\alpha$ in the second cycle quite similar to the first absorption stage suggests that the value at the end of the desorption stage should be regarded as the value for the pure palladium, not the value with slight amount of hydrogen remaining in the bulk. Irreversible changes were also reported previously, such as the change in the shape and the degradation of metallic luster \cite{ProcJpnAcadSerB.85.183}. The observation in this experiment revealed that such a change also happened in microscopic scale.  
Note that the behavior is quite similar for $a_{\beta}$ in the desorption stage, but $a_{\beta}$ went back to its initial value.

The change of $2\theta_\beta$ is large enough to be observed in Fig. \ref{MagAbs}, and detailed time dependent behavior of $a_{\beta}$ is shown in Fig. \ref{AlphaAngle} (a). It had a local minimum during the phase transition, when $t=60$ min, which corresponds to the hydrogen fraction $x \simeq 0.45$ in Fig. \ref{VSH}. These were 0.2\% jump, and possible explanation is the growth of $\beta$ phase first compressed its lattice and then the lattice gradually expanded. The expansion was not all the way back to the initial number. This observation is quite different from the report in Ref. \cite{TransFaradaySoc.58.1989}, where $a_\beta$ increased as the hydrogen fraction $x$ increased. The difference should be due to the experimental condition, particularly if the X-ray diffraction is done during the phase transition or not.

\subsection{Line Width $\Gamma_\alpha$ and $\Gamma_\beta$}
\begin{figure}[!tb]
 \begin{center}
 \includegraphics[width=1.0\columnwidth]{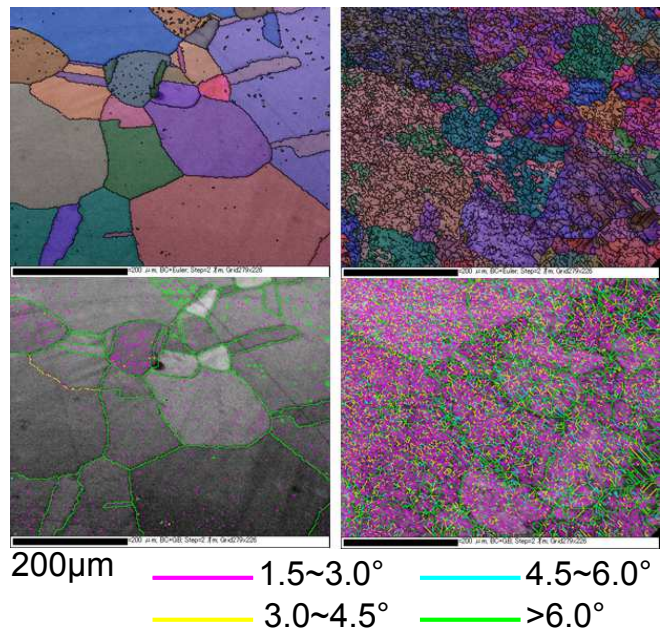}
 \caption{(Color online) SEM-EBSP image of palladium sample before (left) and after (right) hydrogen absorption: top two section shows the crystal orientation, and each color shows different orientation. Bottom two shows the angle difference of crystal orientation at grain boundaries. } 
 \label{SEM}
 \end{center}
\end{figure}

$\Gamma_\alpha$, $\Gamma_\beta$ and the crystal grain size derived from Scherrer formula are summarized in Table \ref{Linewidth}. The trend was that the grain size got smaller as the absorption and the desorption proceeded. The actual grain size in Fig. \ref{SEM} is 10-100 $\mu$m, which is larger than the coherence length of the X-ray, 1000\AA. It is likely that the defect or the disorder in the single crystal grain decreased the effective grain size.  

\begin{table}[!t]
 \caption{Line width and crystal size: the $\alpha$ phase has the value before and after a cycle of absorption and desorption of hydrogen.  The $\beta$ phase value is for after the absorption.} 
\begin{tabular}{lcrclcrcl}
  \hline   \hline 
Phase& &\multicolumn{3}{c}{Line width [deg]} &\hspace{2mm} &\multicolumn{3}{c}{Crystal Size [\AA]}   \\
  \hline   
$\alpha$, before	& &	0.062 & $\pm$ & 0.027	& & 184 & $\pm$ & 91 \\
$\beta$			& &	0.237 & $\pm$ & 0.041	& & 108 & $\pm$ & 14 \\
$\alpha$, after	& &	0.217 & $\pm$ & 0.080	& & 77 & $\pm$ & 19 \\
    \hline   \hline 
\end{tabular}
\label{Linewidth}
\end{table}

The change of $\Gamma_\alpha$ and $\Gamma_\beta$ over time shown in Fig. \ref{AlphaAngle} (b) tells how the reduction of the grain size happened. The $\Gamma_{\alpha}$ increased by 27\% from the beginning after a cycle of absorption and desorption, which was the same trend as the number in Table \ref{Linewidth}. The increase started before the $\alpha$ phase line disappeared. The $\Gamma_\beta$ was approximately the same as $\Gamma_\alpha$ both when the $\alpha$ phase disappeared, and when the $\alpha$ phase reappeared. Once $\alpha$ phase reappeared, $\Gamma_\alpha$ was more or less constant. This means that most increase in $\Gamma$ occurred during the absorption. $\Gamma_\beta$ changed both during the absorption and the desorption, with a local maximum at $t=60$ min for absorption stage and $t=450$ min for desorption stage, though they are not very significant. 

This increase in $\Gamma_\alpha$ is consistent with the observation of crystal orientation with the scanning electron microscope electron backscatter patterns (SEM-EBSP) in Fig. \ref{SEM}. The palladium metal that went through a cycle of absorption and desorption of hydrogen gas had much more small angle tilt grain boundary, between $1.5 ^{\circ}$ and $3 ^{\circ}$, compared to the annealed palladium sample, whereas the size of the grain was roughly the same if the grain boundary is defined with tilt angle larger than $6 ^{\circ}$. The effective decrease of the grain size observed in the X-ray diffraction was due to the formation of small angle tilt grain boundary. This tilt is likely to be formed when the $\alpha$ phase turned into the $\beta$ phase during the absorption. 

\subsection{Diffraction Line Intensity $I_{\alpha}$ and $I_{\beta}$}\label{LineIntensity}
Time dependent behavior of $I_{\alpha}$ and $I_{\beta}$ of the (111) diffraction line is shown in Fig. \ref{AlphaAngle} (c). The $\beta$ phase diffraction line appeared at 8 minutes after the start of the absorption, and two thirds of the change was completed in the first 18 minutes. During this time range, hydrogen fraction $x$ changed from 0.03 to 0.15 as shown in Fig. \ref{VSH}. This $x$ was close to that of when the sudden change in the palladium shape occurred in Ref. \cite{MatSciEngA.551.231}. It is inferred that the sudden change in the shape was induced by the appearance of $\beta$ phase. There was another quick increase of $I_{\beta}$ around $t=60$ min. This corresponds to the hydrogen fraction between $x=0.40$ and $x=0.45$. This jump between $I_{\beta}\simeq 2.1$ and $I_{\beta}\simeq 2.4$ also existed in the desorption side, as the right half of Fig. \ref{AlphaAngle} shows.  

The disappearance of the $\alpha$ phase line slowed down as time went by, but the last moment of the disappearance around $t=60$ min was faster than the earlier part. This jump to $I_\alpha=0$ happened at the same time as the second jump in the $I_\beta$. Also, the fast change in $I_{\alpha}$ and $I_{\beta}$ around $x=0.45$ coincided with the local minimum for $a_{\beta}$ on both the absorption and the desorption stage and the local maximum of $\Gamma_{\beta}$. This whole behavior corresponds to the point of phase transition from the $\alpha+\beta$ phase to $\beta$ phase. Combined with the appearance of $\beta$ phase line, $x$ for $\alpha+\beta$ phase should be $0.05\leq x\leq 0.45$. 

The growth of one phase is faster than the disappearance of the other phase both in the absorption and in the desorption stage.  This breaks the conservation of $I_\alpha+I_\beta$, which should hold if the amount of the material that scatters the X-ray is constant. In our observation, $I_\alpha+I_\beta$ had $\pm 20$\% fluctuation. It is possibly due to the imperfect randomness of the crystal orientation that resulted in the different amount X-ray scattered by the $\alpha$ phase and the $\beta$ phase.  

Most $\beta$ phase diffraction lines totally disappeared when the desorption was completed. This ensures that all the sample went back to $\alpha$ phase and the pumping was long enough to evacuate all hydrogen. 

\subsection{Sample length change $\Delta L$}
$\Delta L$ is plotted in Fig. \ref{AlphaAngle} (d) and Fig. \ref{VSH}. 
The sample expanded by 0.6 \% when it absorbed hydrogen up to PdH$_{0.6}$ and then shrank more than its original length by -2.5\% when hydrogen was released completely. This was roughly consistent with Ref. \cite{ProcJpnAcadSerB.85.183}. The behavior of $\Delta L$ was quite similar to that of $I_\beta$, especially in the desorption stage. Also, the slight change $a_{\alpha}$ and $a_{\beta}$ coincided with the behavior of $\Delta L$, particularly in the desorption stage. These implies the sample length change was related to the phase transition.

\section{Conclusion and Outlook}
To conclude, we performed the {\it in-situ} X-ray diffraction experiment during the phase transition in Pd-H system. The fast growth of $I_\beta$ and gradual decrease in $I_\alpha$ were observed during the absorption of hydrogen. $a_{\alpha}$ and $a_{\beta}$ got smaller as hydrogen concentration got higher with an initial increase in $a_\alpha$ before $\beta$ phase appeared. This suggests the compression of the lattice by the $\beta$ phase. The small angle tilt inside the cyristal grain grew after a cycle of hydrogen absorption and desorption, resulting in the increase of $\Gamma_\alpha$. The coincidence of the behavior in the angle $a_\alpha$, $a_\beta$, $I_{\alpha}$, $I_{\beta}$ and $\Delta L$ implied that the change in the shape was due to the phase transition. 

The reproducibility of the behavior when we have multiple cycles of the absorption and desorption of hydrogen was not clear. The one hour of second cycle suggests that the behavior of parameters were basically the same as the first cycle. However, it is still necessary to directly show what happens when we have more than one cycle of absorption and desorption of hydrogen. 

The behavior of the center of the sample also needs to be investigated. Since X-ray diffraction in principle cannot give the information of the deep inside the bulk, neutron scattering or some other method is expected to reveal how the center of the bulk changes.  

The discussion is only for our observation in Pd-H system. Lots of other systems show superplasticity or similar kind of deformation over a phase transition, same as Pd-H system, but this fact is not enough to show that the microscopic behavior is universal in the phase transition of polycrystalline materials. In order to show the universality, one has to do the same experiment for different materials.

\begin{acknowledgements}
This research is supported by the Strategic Program for R\&D of RIKEN.  The synchrotron radiation experiment were performed at BL44B2 in SPring-8 with the approval of RIKEN (Proposal No. 20100100), and the use of RIKEN beamline at SPring-8 was supported by RIKEN SPring-8 Center. We would like to thank Dr. M. Sato for writing data acquisition. We are grateful to Dr. Nishimura with insightful discussion and Prof. M. Iwasaki and Prof. R. S. Hayano for the stimulating support. 
\end{acknowledgements}

\bibliographystyle{apsrev4-1}
\bibliography{PdH}

\end{document}